\documentclass[aps,prl,twocolumn,showpacs,groupedaddress]{revtex4}  
\usepackage{graphicx}  
\usepackage{dcolumn}   
\usepackage{bm}        
\usepackage{amssymb}   

\newcommand{\met}       {\mbox{$\not\!\!E_T$}}
\newcommand{\vmet}      {\mbox{$\not\!\!\vec{E}_T$}}
\newcommand{\mht}       {\mbox{$\not\!\!H_T$}}

\newcommand{\mlep}       {\mbox{$\not\!\ell$}}
\newcommand{\trkMET}    {\mbox{$\vec{P}_{T}^{trk}$}}

\newcommand{\PYTHIA}    {{\sc{pythia}}}
\newcommand{\ALPGEN}    {{\sc{alpgen}}}
\newcommand{\COMPHEP}   {{\sc{comphep}}}
\newcommand{\GEANT}     {{\sc{geant}}}

\begin{document}
\hspace{5.2in} \mbox{FERMILAB-PUB-06/238-E}
\title{
Search for the Standard Model Higgs Boson
 in the $p\bar{p} \rightarrow ZH \rightarrow \nu \bar{\nu} b\bar{b}$ channel}
%
%
\author{                                                                      
V.M.~Abazov,$^{36}$                                                           
B.~Abbott,$^{76}$                                                             
M.~Abolins,$^{66}$                                                            
B.S.~Acharya,$^{29}$                                                          
M.~Adams,$^{52}$                                                              
T.~Adams,$^{50}$                                                              
M.~Agelou,$^{18}$                                                             
J.-L.~Agram,$^{19}$                                                           
S.H.~Ahn,$^{31}$                                                              
M.~Ahsan,$^{60}$
% ...
\\                                                                            
\vskip 0.30cm                                                                 
\centerline{(D\O\ Collaboration)}                                             
\vskip 0.30cm                                                                 
}                                                                             
\affiliation{                                                                 
\centerline{$^{1}$Universidad de Buenos Aires, Buenos Aires, Argentina}       
\centerline{$^{2}$LAFEX, Centro Brasileiro de Pesquisas F{\'\i}sicas,         
                  Rio de Janeiro, Brazil}                                     
\centerline{$^{3}$Universidade do Estado do Rio de Janeiro,                   
                  Rio de Janeiro, Brazil}                                     
\centerline{$^{4}$Instituto de F\'{\i}sica Te\'orica, Universidade            
                  Estadual Paulista, S\~ao Paulo, Brazil}                     
\centerline{$^{5}$University of Alberta, Edmonton, Alberta, Canada,           
                  Simon Fraser University, Burnaby, British Columbia, Canada,}
\centerline{York University, Toronto, Ontario, Canada, and                    
                  McGill University, Montreal, Quebec, Canada}                
\centerline{$^{6}$Institute of High Energy Physics, Beijing,                  
                  People's Republic of China}                                 
\centerline{$^{7}$University of Science and Technology of China, Hefei,       
                  People's Republic of China}                                 
\centerline{$^{8}$Universidad de los Andes, Bogot\'{a}, Colombia}             
\centerline{$^{9}$Center for Particle Physics, Charles University,            
                  Prague, Czech Republic}                                     
\centerline{$^{10}$Czech Technical University, Prague, Czech Republic}        
\centerline{$^{11}$Center for Particle Physics, Institute of Physics,         
                   Academy of Sciences of the Czech Republic,                 
                   Prague, Czech Republic}                                    
\centerline{$^{12}$Universidad San Francisco de Quito, Quito, Ecuador}        
\centerline{$^{13}$Laboratoire de Physique Corpusculaire, IN2P3-CNRS,         
                   Universit\'e Blaise Pascal, Clermont-Ferrand, France}      
\centerline{$^{14}$Laboratoire de Physique Subatomique et de Cosmologie,      
                   IN2P3-CNRS, Universite de Grenoble 1, Grenoble, France}    
\centerline{$^{15}$CPPM, IN2P3-CNRS, Universit\'e de la M\'editerran\'ee,     
                   Marseille, France}                                         
\centerline{$^{16}$IN2P3-CNRS, Laboratoire de l'Acc\'el\'erateur              
                   Lin\'eaire, Orsay, France}                                 
\centerline{$^{17}$LPNHE, IN2P3-CNRS, Universit\'es Paris VI and VII,         
                   Paris, France}                                             
\centerline{$^{18}$DAPNIA/Service de Physique des Particules, CEA, Saclay,    
                   France}                                                    
\centerline{$^{19}$IPHC, IN2P3-CNRS, Universit\'e Louis Pasteur, Strasbourg,  
                    France, and Universit\'e de Haute Alsace,                 
                    Mulhouse, France}                                         
\centerline{$^{20}$Institut de Physique Nucl\'eaire de Lyon, IN2P3-CNRS,      
                   Universit\'e Claude Bernard, Villeurbanne, France}         
\centerline{$^{21}$III. Physikalisches Institut A, RWTH Aachen,               
                   Aachen, Germany}                                           
\centerline{$^{22}$Physikalisches Institut, Universit{\"a}t Bonn,             
                   Bonn, Germany}                                             
\centerline{$^{23}$Physikalisches Institut, Universit{\"a}t Freiburg,         
                   Freiburg, Germany}                                         
\centerline{$^{24}$Institut f{\"u}r Physik, Universit{\"a}t Mainz,            
                   Mainz, Germany}                                            
\centerline{$^{25}$Ludwig-Maximilians-Universit{\"a}t M{\"u}nchen,            
                   M{\"u}nchen, Germany}                                      
\centerline{$^{26}$Fachbereich Physik, University of Wuppertal,               
                   Wuppertal, Germany}                                        
\centerline{$^{27}$Panjab University, Chandigarh, India}                      
\centerline{$^{28}$Delhi University, Delhi, India}                            
\centerline{$^{29}$Tata Institute of Fundamental Research, Mumbai, India}     
\centerline{$^{30}$University College Dublin, Dublin, Ireland}                
\centerline{$^{31}$Korea Detector Laboratory, Korea University,               
                   Seoul, Korea}                                              
\centerline{$^{32}$SungKyunKwan University, Suwon, Korea}                     
\centerline{$^{33}$CINVESTAV, Mexico City, Mexico}                            
\centerline{$^{34}$FOM-Institute NIKHEF and University of                     
                   Amsterdam/NIKHEF, Amsterdam, The Netherlands}              
\centerline{$^{35}$Radboud University Nijmegen/NIKHEF, Nijmegen, The          
                  Netherlands}                                                
\centerline{$^{36}$Joint Institute for Nuclear Research, Dubna, Russia}       
\centerline{$^{37}$Institute for Theoretical and Experimental Physics,        
                   Moscow, Russia}                                            
\centerline{$^{38}$Moscow State University, Moscow, Russia}                   
\centerline{$^{39}$Institute for High Energy Physics, Protvino, Russia}       
\centerline{$^{40}$Petersburg Nuclear Physics Institute,                      
                   St. Petersburg, Russia}                                    
\centerline{$^{41}$Lund University, Lund, Sweden, Royal Institute of          
                   Technology and Stockholm University, Stockholm,            
                   Sweden, and}                                               
\centerline{Uppsala University, Uppsala, Sweden}                              
\centerline{$^{42}$Physik Institut der Universit{\"a}t Z{\"u}rich,            
                   Z{\"u}rich, Switzerland}                                   
\centerline{$^{43}$Lancaster University, Lancaster, United Kingdom}           
\centerline{$^{44}$Imperial College, London, United Kingdom}                  
\centerline{$^{45}$University of Manchester, Manchester, United Kingdom}      
\centerline{$^{46}$University of Arizona, Tucson, Arizona 85721, USA}         
\centerline{$^{47}$Lawrence Berkeley National Laboratory and University of    
                   California, Berkeley, California 94720, USA}               
\centerline{$^{48}$California State University, Fresno, California 93740, USA}
\centerline{$^{49}$University of California, Riverside, California 92521, USA}
\centerline{$^{50}$Florida State University, Tallahassee, Florida 32306, USA} 
\centerline{$^{51}$Fermi National Accelerator Laboratory,                     
            Batavia, Illinois 60510, USA}                                     
\centerline{$^{52}$University of Illinois at Chicago,                         
            Chicago, Illinois 60607, USA}                                     
\centerline{$^{53}$Northern Illinois University, DeKalb, Illinois 60115, USA} 
\centerline{$^{54}$Northwestern University, Evanston, Illinois 60208, USA}    
\centerline{$^{55}$Indiana University, Bloomington, Indiana 47405, USA}       
\centerline{$^{56}$University of Notre Dame, Notre Dame, Indiana 46556, USA}  
\centerline{$^{57}$Purdue University Calumet, Hammond, Indiana 46323, USA}    
\centerline{$^{58}$Iowa State University, Ames, Iowa 50011, USA}              
\centerline{$^{59}$University of Kansas, Lawrence, Kansas 66045, USA}         
\centerline{$^{60}$Kansas State University, Manhattan, Kansas 66506, USA}     
\centerline{$^{61}$Louisiana Tech University, Ruston, Louisiana 71272, USA}   
\centerline{$^{62}$University of Maryland, College Park, Maryland 20742, USA} 
\centerline{$^{63}$Boston University, Boston, Massachusetts 02215, USA}       
\centerline{$^{64}$Northeastern University, Boston, Massachusetts 02115, USA} 
\centerline{$^{65}$University of Michigan, Ann Arbor, Michigan 48109, USA}    
\centerline{$^{66}$Michigan State University,                                 
            East Lansing, Michigan 48824, USA}                                
\centerline{$^{67}$University of Mississippi,                                 
            University, Mississippi 38677, USA}                               
\centerline{$^{68}$University of Nebraska, Lincoln, Nebraska 68588, USA}      
\centerline{$^{69}$Princeton University, Princeton, New Jersey 08544, USA}    
\centerline{$^{70}$State University of New York, Buffalo, New York 14260, USA}
\centerline{$^{71}$Columbia University, New York, New York 10027, USA}        
\centerline{$^{72}$University of Rochester, Rochester, New York 14627, USA}   
\centerline{$^{73}$State University of New York,                              
            Stony Brook, New York 11794, USA}                                 
\centerline{$^{74}$Brookhaven National Laboratory, Upton, New York 11973, USA}
\centerline{$^{75}$Langston University, Langston, Oklahoma 73050, USA}        
\centerline{$^{76}$University of Oklahoma, Norman, Oklahoma 73019, USA}       
\centerline{$^{77}$Oklahoma State University, Stillwater, Oklahoma 74078, USA}
\centerline{$^{78}$Brown University, Providence, Rhode Island 02912, USA}     
\centerline{$^{79}$University of Texas, Arlington, Texas 76019, USA}          
\centerline{$^{80}$Southern Methodist University, Dallas, Texas 75275, USA}   
\centerline{$^{81}$Rice University, Houston, Texas 77005, USA}                
\centerline{$^{82}$University of Virginia, Charlottesville,                   
            Virginia 22901, USA}                                              
\centerline{$^{83}$University of Washington, Seattle, Washington 98195, USA}  
}                                                                             

\date{July 13$^{th}$, 2006}

\begin{abstract}
We report  a search for the standard model (SM) Higgs boson based on
data collected by the D\O\ experiment at the Fermilab Tevatron Collider,
corresponding to an integrated luminosity of 260~$\mbox{pb}^{-1}$. We study
events with missing transverse energy and two acoplanar $b$-jets, which
provide sensitivity to the $ZH$ production cross section in the $ \nu
\bar{\nu} b \bar{b}$ channel and to $WH$ production, when the lepton from
the $W\rightarrow \ell \nu$ decay is undetected.  The data are consistent
with the SM background expectation, and we set 95\% C.L. upper limits on
$\sigma(p\bar{p} \rightarrow ZH/WH)\times B(H\rightarrow b\bar{b})$ from
3.4/8.3 to 2.5/6.3~pb, for Higgs masses between 105 and 135~GeV.
 

\end{abstract}

\maketitle 

In the standard model
(SM) the Higgs boson~($H$) 
 is responsible for electroweak symmetry breaking and
 has not yet been 
observed. The experiments at the CERN $e^+e^-$ Collider (LEP)
 provide lower limits on its mass, $m_{H} > 114.4$ GeV, while
electroweak global fits favor
a  light Higgs boson,  $m_{H} < 207$ GeV at
95\% C.L.~\cite{LEP}.
If it exists, the Higgs boson could be observed at the Fermilab Tevatron
 Collider (center of mass energy $\sqrt{s}=$1.96~TeV)
by combining different
 analysis channels from both the D\O\ and CDF experiments~\cite{SHWG,Sens}.


We present  a search 
for a SM Higgs boson with $m_H$ 
between 105 and 135~GeV, in the final state with 
missing transverse energy~(\met) and
two or three  jets, 
in which one or two jets are identified (``tagged'') as $b$ jets.
 This final state is sensitive to Higgs bosons produced 
in the $p\bar{p} \rightarrow ZH \rightarrow \nu \bar{\nu} b \bar{b}$ channel, which is 
particularly promising because of
 the expected
large $Z\rightarrow \nu \bar{\nu}$ and $H \rightarrow b\bar{b}$ branching fractions.
The product of cross section ($\sigma$)
 and branching fraction ($B$) is predicted to be about 0.01~pb for a 115 GeV Higgs boson,
which is comparable to that for $WH \rightarrow l \nu b \bar{b}$~\cite{WHZHcross}.

The chosen final state  also has  sensitivity to  $WH$ production 
since the charged lepton from  $W$ decay can be undetected or not identified 
properly ($\mlep \nu b \bar{b}$ channel). 
 Searches for $WH$ production have been performed
previously by relying on the identification of the electron or the muon
from  leptonic $W$ decay~\cite{WH-d0,WH-cdf}.

There are two main sources of background to this final state:
i) the "physics" backgrounds $Z$+jets, $W$+jets, electroweak diboson 
production~($WZ$ and $ZZ$), 
and top quark production with undetected leptons or jets, and
ii) a large instrumental background  caused by multijet 
events with mismeasured jet energies that is difficult to simulate.
In the $ZH$ or $WH$ processes,
since the two $b$ jets are boosted along the Higgs-momentum direction,  
they are  not back-to-back in azimuthal angle ($\varphi$), 
in contrast to the dominant  dijet background.
Our search  is 
based on an integrated luminosity of 260~$\mbox{pb}^{-1}$ accumulated with 
a dedicated trigger designed to select events with 
 significant \met and with
jets that are
not back-to-back.


The D\O\ tracking system, consists of a 
silicon microstrip tracker (SMT) and a central fiber tracker (CFT), 
both located within a 2~T superconducting solenoidal 
magnet~\cite{d0det}, with tracking and 
vertexing at pseudorapidities $|\eta|<3$ and $|\eta|<2.5$, respectively, 
where $\eta = -\ln{(\tan{(\theta /2)})}$, and $\theta$ is the polar angle.
A liquid-argon and uranium calorimeter has a 
central section (CC) covering  $|\eta|$ up to 
$\approx 1.1$, and two end calorimeters (EC) that extend coverage 
to $|\eta|\approx 4.2$~\cite{run1det}. An outer muon system, at $|\eta|<2$, 
consists of a layer of tracking detectors and scintillation trigger 
counters in front of 1.8~T toroids, followed by two similar layers 
after the toroids.

To estimate the number of expected events, the signal ($ZH, WH$), $t\bar{t}$,
 and  diboson production is 
simulated with \PYTHIA~\cite{pythia}. 
For $W$ and $Z$ events with  two  or more jets we use \ALPGEN~\cite{ALPGEN}, 
and for single top simulation we use \COMPHEP~\cite{COMPHEP}. 
The samples generated by \COMPHEP~and \ALPGEN~are
 passed through \PYTHIA~for showering and hadronization.
The cross section for the \ALPGEN\ samples  are
normalized to next-to-leading-order
calculations~\cite{MCFM}.
All the samples are
 processed through  D\O\ detector simulation based on \GEANT~\cite{GEANT},
and  D\O\ reconstruction software.
Trigger efficiencies measured in data are  applied to correct the simulated events.

Event selection requires two or three jets reconsructed with
the "iterative-midpoint-Run-II" cone algorithm, 
with   $p_T>20$~GeV, $|\eta|<2.5$ and a
cone radius of $\Delta R = \sqrt{(\Delta \eta)^2 + 
(\Delta \varphi)^2} <$ 0.5. 
Jets are required to pass quality criteria designed to reject noise and 
suppress electron- or photon-induced energy depositions,
and jet energies are corrected to the 
particle level 
using jet energy calibration and resolution 
factors determined from
photon+jet events. 
Corrections depend on the $p_T$ and 
$\eta$ of the jet and are
 typically 30\%.
Jet energy resolution 
varies from
 20\% to 10\% for $p_T$ between 40 and 150 GeV.
%

The primary vertex  has to be within $\pm 35 $ cm in the $z$
direction, and at
least two ``taggable'' jets passing  the above 
requirements  must be present in the event
to be included in our final sample. A jet is taggable
if it contains within its cone at least 
two tracks satisfying strict 
quality criteria, one with $p_T >$ 1~GeV, and another with $p_T >$ 0.5 GeV.
The average fraction of taggable jets is measured using
  $W(\rightarrow \mu \nu)$+jets
data, and is (86$\pm$1)\% per jet.
This fraction, which is a function of $\eta$ and $p_T$ of the jet, and of the
$z$ coordinate of the primary vertex, is used to correct
the simulated jets.
%
%

We then require: i) \met $>$ 50~GeV, where
\met\ is calculated from the position and energy of the calorimeter cells,
ii)
the azimuthal angle between the 
two  highest $p_T$
(leading) jets to be less than 165$^{\circ}$, and iii)
no isolated electrons or muons, 
in order  to suppress multijet background and
$W(\rightarrow e\nu, \mu\nu)$+jet, and $Z(\rightarrow ee, \mu\mu)$+jet events.
For the rejection 
of $t\bar{t}$ background, we require 
the scalar sum $H_T$ of the $p_T$ of the jets to be less than 240~GeV.
To further reduce instrumental background induced by mismeasurement of jet energy,
which produces abnormal \met , we define 
$min\Delta \varphi$(\vmet , jets) as 
the minimum  difference in  $\varphi$ between the direction of \vmet\
 and any of the jets, 
\mht $\equiv$ $|\sum_{i=1}^{njet}{\vec{p}_T}|$ as 
the magnitude of the vector sum of the  $\vec{p}_T$ of the jets, 
$\vec{P}_{T}^{trk}$ $\equiv$ $-\sum_{i=1}^{ntrk}{\vec{p}_T}$ as  
opposite vector sum of the  $\vec{p}_T$ of all tracks,
$\Delta \varphi$(\vmet , \trkMET) as 
the difference in $\varphi$ between the direction of \vmet\ and \trkMET , and
A(\met, \mht)$\equiv$(\met $-$ \mht)/(\met $+$ \mht) as 
the asymmetry between \met\ and \mht.
The instrumental background is significantly reduced by requiring:
 \met\ (in GeV) $> 80-40 \times min\Delta \varphi$(\vmet, jet), \ \
 $|\trkMET | >  20$~GeV, \ \ $\Delta \varphi$(\vmet, \trkMET) 
$<  
\frac{\pi}{2}$\ \,
 and \ $-$0.1 $<$ A(\met, \mht) $<$ 0.2. All these requirements define the signal
region.

$W(\rightarrow \mu \nu)+$jets data are 
used to confirm that the  above variables  are well modeled. 
The instrumental background is then estimated from the data using the signal 
and a ``sideband''
region, which is defined by  requiring all  above selections, except for the 
 requirement 
$\Delta \varphi$(\vmet , $\trkMET ) > \frac{\pi}{2}$.
The distribution in the simulated instrumental background generated by \PYTHIA~ gives a 
qualitative description of this background.
This indicates that we are correctly identifying the
background source, and we therefore model it using sideband data to avoid
uncertainty from the difficult simulation of instrumental background. 
The physics backgrounds passing  the final selection tend to be distributed around 
$\Delta \varphi (\vmet, $\trkMET $) \sim 0$, while the instrumental background 
is distributed similarly in the sideband and in the signal region due to 
mis-measurement of  
jet energy or of charged tracks. 

\begin{figure}
\includegraphics[scale=0.4]{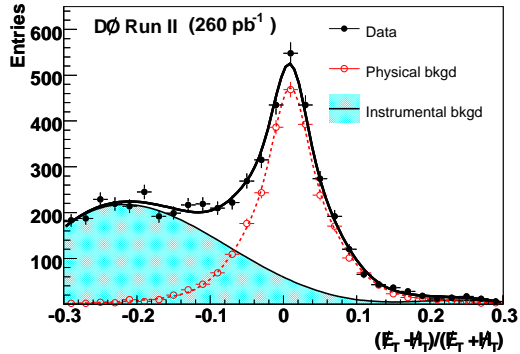}
\caption{\label{fig:inst} Asymmetry distribution A(\met, \mht)  in the signal region, 
prior to the imposition of 
 the requirement on  A(\met, \mht). The data is described by 
the sum of the physics background, modeled by a triple Gaussian, and
 the instrumental background modeled by a polynomial function.
}
\end{figure}

\begin{figure}
\includegraphics[scale=0.37]{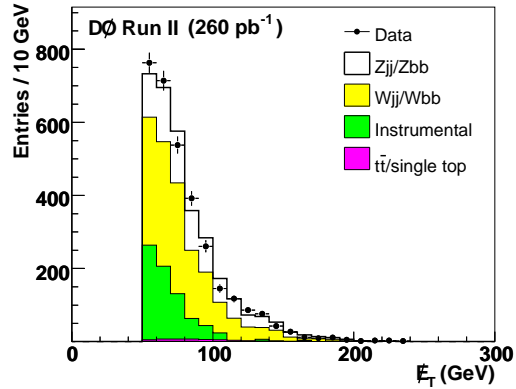}
\caption{\label{fig:met} \met\ distribution after  selection except for  $b$-tagging.
}
\end{figure}

Figure~\ref{fig:inst} shows the $A(\met, \mht)$ distribution in the
 signal region.
The amount of physics background in the signal region  is estimated using
the simulation, and parameterized by a triple Gaussian (TG) function, shown as a dashed line 
in Fig.~\ref{fig:inst}.
The contribution not described by this parameterization is considered to be
the instrumental background, and is modeled with a polynomial function 
tested with a fit to the data in the sideband region.
The physics
background contributes about 15\% of the events in the sideband
region and is included in the model of instrumental background.
%
The sum of the absolutely normalized TG parameterization
and of the  polynomial function  is then
  fitted to the data in the signal region, as shown in Fig.~\ref{fig:inst}.
(Before $b$-tagging,
the Higgs signal is negligible.)
The instrumental background in the signal region  amounts
to 696$\pm$91~events, while the 
 physics background amounts to  2520$\pm$330 events.
%
%
Since our search requires good modeling of \met , we show in
Fig.~\ref{fig:met} 
  the \met\ distribution after all 
 requirements, excepting $b$-tagging.  The data are well described by the
sum of the simulation of $Z/W+jj/b\bar{b}$ and
 the estimated contribution from
 instrumental background.  Top pair and single top pair production
 represent negligible  contributions before requiring $b$-tagging.
%
%

To select $b$ jets, we apply a 
 $b$-tagging algorithm that uses
a jet lifetime probability (JLIP) computed from the tracks 
associated with the jet.
A small probability corresponds to jets having tracks with a large 
impact parameter that characterize $b$-hadron decay. 
We use two samples for our search: one that requires the two leading
jets to pass the $b$-tagging condition 
(double $b$-tagged sample, or DT sample); the 
other requires exactly one jet to pass the $b$-tagging condition,
and does not accept events from the DT sample
(exclusive single $b$-tagged sample, or ST sample). 
The requirements on the lifetime probability are defined by optimizing 
 the sensitivity
to Higgs signal.
In the DT sample, we require JLIP $< 1\%$ for the leading jet
 and $< 4~\%$ for the second-leading jet. 
In the ST sample we require a more stringent JLIP $ < 0.1~\%$.
The average 
$b$-tagging efficiency is $\approx$ 50\% (40\%, 30\%)  for JLIP $< 4\% 
(1\%, 0.1\%)$.
  The relative uncertainty on the $b$-tagging efficiency is  7\% per jet.
The mis-tag rate is defined as the fraction of light-quark jets tagged as 
$b$ jets, and its average value is
approximately the  value of the JLIP requirement.
For the instrumental background, we estimate the mis-tag rate from  data in the sideband region,
and extrapolate it into the signal region.
Table~\ref{tab:table1} lists the number of $ZH$ and $WH$ signal, background and observed 
events for each 
$b$-tag requirement, and also for the inclusive 
sample of events with at least one $b$-tagged jet
with  JLIP $< 4\%$ (to verify that the data are also well described by the simulation 
in  another $b$-tagging configuration).
After the ST requirement, 106 events remain, while $94.5 \pm 17.0$  events are expected.
In the DT sample, we observe 25 events, while $27.0 \pm 5.1$ are expected, and in the 
inclusive sample these numbers are 592 and $555 \pm 70$ events, respectively.

\begin{table}
\caption{\label{tab:table1} Number of expected  signal (for $m_H=115$ GeV), background,
and observed events (obs.) before 
$b$-tagging, after inclusive (IST), and exclusive (ST) single $b$-tagging, and after double
$b$-tagging (DT).
%
Before $b$-tagging, the expected background is by construction
equal to the observed events (see text on the background determination).
%
%
%
The numbers of events after the $\pm 1.5 $ standard deviation (s.d.) 
mass window  requirement are given in parenthesis. 
The errors on these numbers are in average 18\% (19\%) for the
ST (DT)  sample.
%
%
}
\begin{ruledtabular}
\begin{tabular}{ccccc}
              & \met\ +      & \met\ +      & \met\ +            & \met\ +      \\
              & 2, 3 jets & 2, 3 jets & 2, 3 jets       & 2, 3 jets     \\
              &      &   IST & ST  & DT       \\ \hline
$ZH$                   & 0.71     & 0.62          & 0.26 (0.20)           & 0.24 (0.21)        \\
$WH$                   & 0.54     & 0.47          & 0.20 (0.15)           & 0.18 (0.15)        \\
$Zjj$                  & 843      & 93.3          & 7.9  (2.6)            & 1.4 (0.5)           \\
$Wjj$                  & 1600     & 260         & 36.1 (13.6)           & 4.2 (1.5)          \\ 
$Zbb$                  & 13.1     & 11.3          & 4.7  (1.6)            & 4.1 (1.4)          \\
$Wbb$                  & 12.4     & 10.5          & 4.4  (1.4)            & 3.6 (1.1)          \\ 
$t\bar{t}$/$tb$/$tqb$  & 42.3     & 33.6          & 15.3 (5.6)            & 9.0 (3.0)           \\
$WZ$/$ZZ$              & 7.3      & 3.4           & 1.1  (0.71)           & 0.9 (0.6)           \\ 
Instrumental     & 696      & 143          & 25.0 (8.4)            & 3.9 (1.3)            \\
Total expectation      & $\equiv$ obs.     & 555         & 94.5 (34.0)           & 27.0 (9.4)          \\
Observed events        & 3210     & 592           & 106 (33)              & 25 (11)              \\
\end{tabular}
\end{ruledtabular}
\end{table}


We estimate the systematic uncertainty due to trigger  and jet reconstruction 
efficiency, jet energy calibration, jet resolution, $b$-tagging, 
instrumental-background estimation, physics-background cross sections and parton distribution 
functions, by varying each source of uncertainty by $\pm 1 $ s.d. and 
repeating the analysis. 
The systematic uncertainties are estimated  separately for the DT and ST samples.
In total, we find a 19\%~(14\%) uncertainty on  signal acceptance and 
19\%~(18\%) uncertainty on the total background for the DT~(ST) analysis. The dominant
systematic uncertainties are due to $b$-tagging and jet reconstruction and calibration.
The uncertainty on the integrated luminosity is  6.5\%.

\begin{figure}
\includegraphics[scale=0.37]{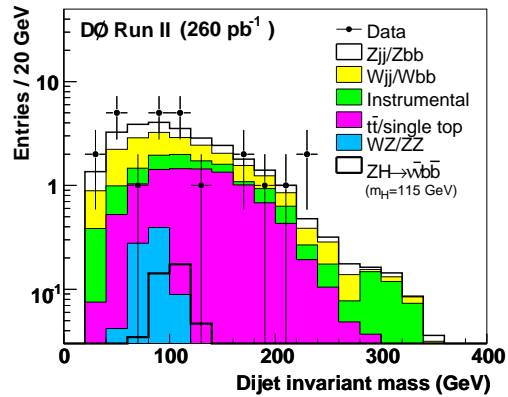}
\caption{\label{fig:dijet} Dijet invariant mass distribution in the DT sample. The expectation originating from $ZH$
production with $m_H=115$ GeV is also shown.}
\end{figure}

\begin{figure}
\includegraphics[scale=0.4]{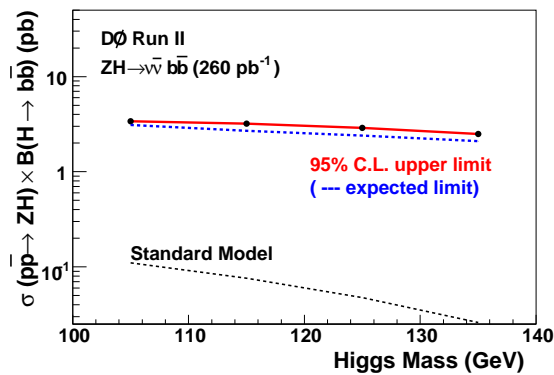}
\caption{\label{fig:limit} 95\% C.L. upper limit on 
 $\sigma (p\bar{p} \rightarrow ZH) \times B(H \rightarrow b \bar{b})$
(and corresponding expected limit)
for  $ZH$ production vs. Higgs mass, as derived from the ST+DT combination.
}
\end{figure}

We then search for an excess of events as a function of $m_H$
 by counting events in the dijet mass distribution within a  $\pm 1.5$ s.d.
window around the reconstructed 
Higgs-boson mass peak, e.g., $\pm 25.2 $ GeV 
for $m_H$=115 GeV.
No excess over the SM background is found in the data,
as can be seen for the DT dijet mass distribution in Fig.~\ref{fig:dijet}, 
in which the expected $ZH$ signal for $m_H =115 $ GeV is also shown.
The acceptance for $ZH$ ($WH$) events is 1.04\% (0.43\%) for $m_H= 115$ GeV.
 We thus
set  95\% C.L. upper
 limits on
$\sigma(p\bar{p}\rightarrow ZH)\times B(H\rightarrow b\bar{b})$ and 
$\sigma(p\bar{p}\rightarrow WH)\times B(H\rightarrow b\bar{b})$, 
 using 
a modified frequentist approach, the $CL_S$ method~\cite{cls}.
In this method, the binned distributions are summed 
over the log-likelihood ratio test statistic. 
Systematic uncertainties are incorporated 
into the signal and background expectations using Gaussian sampling of individual uncertainties. 
For the  limits obtained when combining the likelihoods of the ST and DT analyses,
 correlations between uncertainties are handled by 
varying simultaneously all identical 
sources. Limits are determined 
by scaling the signal expectations until the
 probability for the background-only hypothesis falls below 5\% (95\% C.L.).
This translates into a cross-section limit 
for $\sigma(p\bar{p}\rightarrow ZH)\times B(H\rightarrow b\bar{b})$ of 
3.2~pb and for $\sigma(p\bar{p}\rightarrow WH)\times B(H\rightarrow b\bar{b})$ of 7.5~pb, 
assuming $m_H=$ 115~GeV.
The limits for four Higgs mass points~(105,
 115,
 125, and 135~GeV) and for ST, DT, and the combined ST+DT results
 are summarized in Tables~\ref{tab:table2} 
and~\ref{tab:table3}.
We set 95\% C.L.  upper limits from 3.4 to 2.5~pb on 
$\sigma(p\bar{p}\rightarrow ZH)\times B(H\rightarrow b\bar{b})$ 
for  $m_H =$ 105--135 GeV (Fig.~\ref{fig:limit}).
%
The CDF collaboration has
published  combined limits (ST+DT)  with Tevatron Run I data, i.e.
at $\sqrt{s}=1.8$ TeV,   of 7.8--7.4 pb
for $m_H=$ 110--130 GeV~\cite{cdf-runI}.

In conclusion, we have performed a search for $ZH$ and $WH$ associated production 
in the \met + $b$ jets channel 
using 260 pb$^{-1}$ of data.
We have studied the dijet mass spectrum of the two leading jets 
with double and exclusive single $b$-tagged jets for 
Higgs boson masses between 105 and 135~GeV.
In the absence of signal,
 we have set upper limits on different Higgs production channels/final states,
 and have combined them. The  combined limits are
between 3.4 to 2.5~pb (8.3 to 6.3~pb) on the cross section for 
$ZH$~($WH$) production multiplied by the branching fraction for $H\rightarrow b\bar{b}$.
These are the first limits in the $ZH$ channel based on Tevatron Run II data.

\begin{table}
\caption{\label{tab:table2} Expected/observed 95\% C.L. limits on 
 $\sigma (p\bar{p} \rightarrow ZH) \times B(H \rightarrow b \bar{b})$
in pb, as a function of $m_H$.}
\begin{ruledtabular}
\begin{tabular}{ccccc}
Higgs mass (GeV) & 105 & 115 & 125 & 135 \\ \hline
ST  & 7.7/8.2 & 6.8/6.8 & 6.0/7.3 & 5.4/7.5 \\
DT  & 3.3/4.2 & 2.8/3.6 & 2.5/2.8 & 2.2/2.2 \\
ST+DT & 3.1/3.4 & 2.7/3.2 & 2.4/2.9 & 2.1/2.5 \\
\end{tabular}
\end{ruledtabular}
\end{table}
\begin{table}
\caption{\label{tab:table3} Expected/observed 95\% C.L. limits on
 $\sigma (p\bar{p} \rightarrow WH) \times B(H \rightarrow b \bar{b})$ 
in pb, as a function of $m_H$.}
\begin{ruledtabular}
\begin{tabular}{ccccc}
Higgs mass (GeV) & 105 & 115 & 125 & 135 \\ \hline
ST   & 18.5/17.6 & 15.9/16.9 & 14.9/18.9 & 12.4/18.5 \\
DT  & 8.0/9.6 & 6.6/8.1 & 6.3/7.1 & 5.3/5.3 \\
ST+DT  & 7.6/8.3 & 6.3/7.5 & 6.0/7.4 & 5.0/6.3 \\
\end{tabular}
\end{ruledtabular}
\end{table}

%
We thank the staffs at Fermilab and collaborating institutions, 
and acknowledge support from the 
DOE and NSF (USA);
CEA and CNRS/IN2P3 (France);
FASI, Rosatom and RFBR (Russia);
CAPES, CNPq, FAPERJ, FAPESP and FUNDUNESP (Brazil);
DAE and DST (India);
Colciencias (Colombia);
CONACyT (Mexico);
KRF and KOSEF (Korea);
CONICET and UBACyT (Argentina);
FOM (The Netherlands);
PPARC (United Kingdom);
MSMT (Czech Republic);
CRC Program, CFI, NSERC and WestGrid Project (Canada);
BMBF and DFG (Germany);
SFI (Ireland);
The Swedish Research Council (Sweden);
Research Corporation;
Alexander von Humboldt Foundation;
and the Marie Curie Program.

\end{document}